\documentclass[twocolumn,preprintnumbers,endnote,nofootinbib,prd,9pt,superscriptaddress]{revtex4}
\pdfoutput=1

\usepackage[utf8]{inputenc}
\usepackage{graphicx}
\usepackage{amssymb}
\usepackage{amsxtra}
\usepackage{amsmath}
\usepackage{booktabs,multirow,tabularx}
\usepackage{slashed}
\usepackage{float}
\usepackage{placeins}
\usepackage{rotating}
\usepackage{lscape}
\usepackage{color}
\usepackage{hyperref}

\usepackage[Q=yes,pverb-linebreak=no]{examplep}

\DeclareGraphicsExtensions{.pdf}
\graphicspath{{./figures/}}

\newcommand{\vev}[1]{\langle {#1} \rangle}

\newcommand{\ord}[1]{\mathcal{O}{(#1)}}
\newcommand{\eq}[1]{Eq.~(\ref{#1})}

\def\beq{\begin{equation}}
\def\bea{\begin{eqnarray}}
\def\eeq{\end{equation}}
\def\eea{\end{eqnarray}}
\def\beqnl{\begin{align}}
\def\endal{\end{align}}

\begin{document}

\title{\boldmath 
Electron $g-2$ Foreshadowing Discoveries at FCC-ee
}

\author{Hooman Davoudiasl\footnote{email: hooman@rcf.rhic.bnl.gov}}

\affiliation{High Energy Theory Group, Physics Department, Brookhaven National Laboratory, Upton, NY 11973, USA}

\author{Pier Paolo Giardino\footnote{email: pierpaolo.giardino@usc.es }}

\affiliation{Instituto Galego de F\'{i}sica de Altas Enerx\'{i}as, Universidade de Santiago de Compostela,\\
15782 Santiago de Compostela, Galicia, Spain}

\begin{abstract}

A future $e^+e^-$ circular collider (FCC-ee) may provide a unique probe of the electron Yukawa coupling through  Higgs boson production on resonance. 
Motivated by this exciting  possibility,   we examine a simple model which can result in  $\ord{10}$ modifications of  the Higgs coupling to electrons.  The  model can also lead to  deviations in the electron anomalous magnetic moment, $g_e-2$, which at present shows a $+2.2\sigma$ or $-3.7\sigma$ deviation, implied by  differing precision determinations of the electromagnetic  fine structure constant.
 The electron $g_e-2$ can be a forerunner for FCC-ee discoveries which, as we elucidate, may not be accessible to the high-luminosity LHC measurements.  A simple extension of our model can also account for the current  deviation in the muon $g_\mu-2$.         

\end{abstract}

\maketitle

\section{Introduction\label{sec:intro}}

Open questions that are not answered within the Standard Model (SM), such as the identity  of dark matter and the origin of the cosmic baryon asymmetry, highlight the need for new fundamental  physics to describe the observed phenomena in Nature.  It is often assumed that the new physics may manifest itself through its interactions with the Higgs field.  As such, deviations in the properties of the third generation fermions are deemed most likely, due to their stronger coupling with the Higgs and hence the fundamental new physics underlying its potential.

Despite the above expectations, one may entertain the possibility that new physics may show up in the Higgs couplings  of the first generation fermions.  In particular, the electron can provide clean signals, unlike the top or the tau.  In the SM, the Yukawa coupling of the electron to the Higgs is extremely small, $\ord{10^{-6}}$, and very challenging to measure directly.  However, this can provide an opportunity to find new effects unambiguously, far from the SM expectations.

In this work,
 we propose a simple model which can lead to $\ord{10}$ enhancement of the electron Yukawa coupling.  The model, in its basic form, only requires a new vector-like lepton and a weak scale singlet scalar.  The SM Higgs coupling to electrons can be potentially probed at a future $e^+e^-$ circular collider (FCC-ee) \cite{dEnterria:2017dac}, down to $\sim 1.6$ times the SM value \cite{dEnterria:2021xij}, which would be an impressive improvement over the current limit at $\sim 260$ times the SM expectation \cite{CMS:2014dqm,ATLAS:2019old}.  This bound is projected to be improved to $\sim 120$ times the SM value by the end of the high-luminosity LHC (HL-LHC) running \cite{Bernardi:2022hny}, far from the potential projection for FCC-ee, using resonant production.  
Hence, the model we propose can be tested at the FCC-ee.  It is interesting that the type of physics we postulate may not be detectable by the HL-LHC, with $\sim 3$~ab$^{-1}$ of integrated luminosity.  However, quite generally, we expect that the model can lead to deviations in the anomalous magnetic moment of the electron $g_e-2$, at levels that could be accessible to  experiment in the coming years.  With a modest extension, the model can also address the deviation in the muon $g_\mu-2$ \cite{Muong-2:2006rrc,Aoyama:2020ynm,Borsanyi:2020mff,Muong-2:2021ojo}, which is still under investigation by theory and experiment.

Ideas for measuring the Yukawa couplings of first generation fermions, using  atomic clocks,  have been considered in Ref.~\cite{Delaunay:2016brc}. 
Enhanced electron Yukawa coupling in the context of 2-Higgs doublet models has been discussed in Ref.~\cite{Dery:2017axi}.  A model of charged lepton masses which can lead to deviations in the electron and muon Yukawa couplings and their values of $g-2$ is presented in Ref.~\cite{Chang:2022pue}.  In Ref.~\cite{Dermisek:2023tgq}, connections between the muon dipole  moments and its Yukawa coupling have been examined. 


\section{\boldmath Electron $g-2$}

Precision measurements of the electron and muon $g-2$ can provide stringent tests of the SM.  Currently, the status of $a_e\equiv (g_e-2)/2$ is not clear, since the most precise measured values of the fine structure constant $\alpha$ do not agree.  The most recent experimental value obtained using rubidium (Rb) atoms is \cite{Morel:2020dww}
\beq
\alpha^{-1}({\rm Rb}) = 137.035999206(11)\,,
\label{alphaRb}    
\eeq
which disagrees with that obtained earlier by another group using cesium (Cs) atoms \cite{Parker:2018vye} 
\beq
\alpha^{-1}({\rm Cs}) = 137.035999046(27)\,,
\label{alphaCs}    
\eeq
leading to a discrepancy at the level of $5.5\,\sigma$.

The SM prediction for $a_e^{\rm SM}$ from Ref.~\cite{Aoyama:2017uqe}, when compared to the latest experimental determination   $a_e^{\rm exp}$  \cite{Fan:2022eto},  results in a deviation $\Delta a_e \equiv a_e^{\rm exp} - a_e^{\rm SM}$, which depends on the value of $\alpha$ used as input.  For $\alpha({\rm Rb)}$, we find    
\bea
\Delta a_e({\rm Rb})\!\!&\equiv&\!\! 
a_e^{\rm exp} - a_e^{\rm SM}({\rm Rb})\\ \nonumber
&=& \!\![34 \pm 13({\rm exp}) \pm 9(\alpha) \pm 2 {\rm (th)]}\times 10^{-14}.
\label{Del-ae-Rb}
\eea
Summing errors in quadrature, 
we get 
\beq
\Delta a_e ({\rm Rb}) = (34 \pm 16)
\times 10^{-14}\,,
\label{Del-ae-Rb-quad}
\eeq
which leads to a positive deviation of $2.2\sigma$. 
However, using the value $\alpha({\rm Cs})$ yields   
\bea
\Delta a_e ({\rm Cs})\!\! &\equiv&\!\! 
a_e^{\rm exp} - a_e^{\rm SM}({\rm Cs})\\ \nonumber
&=& \!\![-101 \pm 13({\rm exp}) \pm 23(\alpha) \pm 2 {\rm (th)]}\times 10^{-14},
\label{Del-ae-Cs}
\eea
which gives 
\beq
\Delta a_e ({\rm Cs})= (-101 \pm 27)
\times 10^{-14} 
\label{Del-ae-Cs-quad}
\eeq
and thus leads to a negative  discrepancy of $3.7 \sigma$.\footnote{A deviation of $|\Delta a_e|\sim 8\times 10^{-14}$, within $\sim 1.6\sigma$ from $\Delta a_e(\rm Rb)$, can be achieved using only SM fields by increasing the Yukawa coupling between the Higgs and the electron to $\sim 250$ times its SM value, which is very close to the current experimental bounds.  The deviation  $\Delta a_e(\rm Cs)$ can be addressed in this way only within $\sim 3.4\sigma$.}  

We note that the discrepancy between theory and experiment has grown since the experimental determinations of $\alpha$ in 2018 \cite{Parker:2018vye} and 2020 \cite{Morel:2020dww}, for either value used as input.  This is due to the new experimental result for $a_e^{\rm exp}$ \cite{Fan:2022eto}, which is smaller than the previous determination \cite{Hanneke:2008tm} by $14 \times 10^{-14}$, but has less than half the uncertainty of the earlier measurement.

\section{The Model}

We consider a theory where we add to the usual SM field content a singlet scalar $\phi$ and a family of heavy vector-like fermions $S_l$. The fermion $S_l$, having the  gauge charges of the SM right-handed electron,  carries lepton flavor number.   Therefore, we have three distinct fermions $S_e$, $S_\mu$, $S_\tau$. The Lagrangian of this model is given by 
\bea
\mathcal{L}&=&\mathcal{L}_{\rm SM} + \mathcal{L}_{\phi} + \mathcal{L}_{S_e} + \mathcal{L}_{S_\mu} + \mathcal{L}_{S_\tau}+ \text{\small H.C.}\label{TotLagrangian}\\
\mathcal{L}_{\phi}&=&\frac12\partial_\mu\phi\,\partial^\mu\phi-\frac12 m_\phi^2\phi^2\nonumber\\&& -\frac{\mu}{3!}\, \phi^3 - \frac{\lambda}{4!}\, \phi^4 - \kappa\, v\, \phi H^\dagger H
\label{Lphi}\\
\mathcal{L}_{S_e}&=& i\bar{S}_e  D\!\!\!\!/ \,S_e- M_{S_e} \bar{S}_e S_e- M_{e,S_e} \bar{S}_{e,L} e_R -y_{S_e} \bar{L_e}H S_{e,R}\nonumber\\&& -\xi_e \phi \bar{S}_{e,L} e_R-g_{S_e} \phi \bar{S}_{e} S_{e} + \small{\rm H.C.}
\label{LagrangianS}\\
 {\cal L}_{S_\ell}&=&
 {\cal L}_{S_e}(e\to\ell); \quad \ell=\mu,\tau,
\eea
where $H$ is the usual Higgs doublet, and $v$ is the SM Higgs vacuum expectation value: $\vev{H}= v/\sqrt{2} \approx 174$~GeV. As explained in the introduction, in this paper we are interested particularly in the modification of the physics of the electron, therefore for now we are going to focus only on the first three terms of the Lagrangian in Eq. (\ref{TotLagrangian}). We will come back to the other terms at the end of the paper. 

It has to be noted that the above setup, as it stands could lead to charged lepton flavor violation.  The form of the interactions in \eq{LagrangianS} implicitly assumes that such effects are absent or else sufficiently small, which is a phenomenologically motivated choice.    However, without a good symmetry, one can in principle couple $S_e$ to all three SM lepton generations and end up with flavor off-diagonal interactions.  These could provide additional signals for our model, but one has to make sure they are not at unacceptable levels.  This is a generic model building problem, whenever new vector-like leptons are introduced. We will address this issue in the Appendix.

Note that while at the level of the unbroken symmetry $\phi$ cannot mix directly with the doublet field $H$, once the electroweak symmetry is broken $\phi$ can mix with the Higgs boson $h$ (corresponding to the observed scalar at $\sim 125$~GeV) and new mass mixing terms between $S_e$ and the electron appear. Consequently, both the mass of the electron $m_e$ and the effective coupling $y_e^h/\sqrt{2}$ between the Higgs boson $h$ and the electron are modified. However, since the modifications to these parameters depend in different ways on the Lagrangian parameters, the SM relation between $m_e$ and $y_e^h$ is not preserved. This is the central mechanism which allowed us to obtain a large $y_e^h$, while keeping $m_e$ at its measured value. 
 We will illustrate this point using an effective field theory (EFT) approach, next.  

\section{Effective Field Theory Analysis}
In order to elucidate the above mechanism, we now provide a simplified EFT analysis.  To do so, let us assume parameters such that the cubic and the quartic terms in the $\phi$ potential can be ignored, focusing only on the mass term.\footnote{The last term in Eq. (\ref{Lphi}) generates a tadpole term for $\phi$, and therefore a vacuum expectation value ($v_\phi$).  In our analysis, we have implicitly assumed the value of $v_\phi$ to be small compared to $m_\phi$. We find that, in the decoupling limit $\kappa\to0$, the potential of $\phi$ has only one minimum around the origin if $\mu< \sqrt{3\lambda}\, m_\phi$. Under this condition, which remains approximately true in the full theory as long as $\kappa\ll 1$, the resulting $v_\phi$ is small and mostly independent of  the exact values of $\lambda$ and $\mu$}. Treating both the $S_e$ fermion and $\phi$ as heavy fields, we may integrate them out and obtain the following dimension-6 operator: 
\beq
O_6 = \kappa\frac{(H^\dagger H) {\bar L} H e_R}{\Lambda^2}\,,
\label{O6}
\eeq
where 
\beq
\Lambda^2 \equiv y_{S_e}^{-1} \xi_e^{-1} v^{-1} M_{S_e} m_\phi^2\,,
\label{Lambda}
\eeq 
using the notation from \eq{LagrangianS}.

After electroweak symmetry breaking, the contribution from the usual dimension-4 Higgs Yukawa coupling to the electron, $y_e H\bar L e_R$, and that from the $O_6$ of the EFT yield the following for the electron mass
\beq
m_e = \frac{y_e\, v}{\sqrt{2}} + \frac{\kappa\, v^3}{\sqrt{8}\,\Lambda^2}.
\label{EFT-me}
\eeq
The corresponding effective coupling $(y_e^h/\sqrt{2}) h \, \bar e e$ of the electron to $h$ will then be given by
\beq
y_e^h = y_e + 
\frac{3 \kappa v^2}
{2 \,\Lambda^2}.
\label{EFT-yeh}
\eeq
From \eq{EFT-me}, we see that one can choose $y_e$ to be much larger than the SM value, as long as there is a sufficient degree of cancellation between the two terms that contribute to the electron mass, resulting in the measured value $m_e \approx 0.511$~MeV.  However, such a cancellation will no longer be maintained for $y_e^h$, leading to a value $y_e^h \sim \ord{y_e}$, which can be much larger than $y_e^{h,\rm SM}=\sqrt{2} \, m_e/v \approx 3\times 10^{-6}$ predicted in the SM.

\section{Lepton Magnetic Dipole Moments}

\begin{figure}[t]
	\centering
\includegraphics[width=\columnwidth]{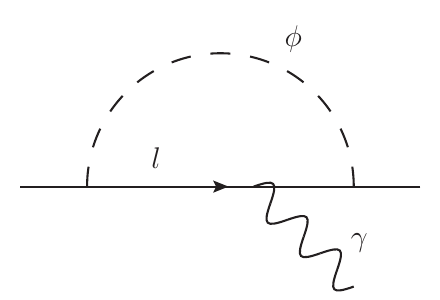}  
	\caption{Contribution to the $g-2$ of a lepton $l$ induced by its coupling to the scalar. $\phi$}
	\label{g2e}
\end{figure}

\begin{figure}[t]
	\centering
\includegraphics[width=\columnwidth]{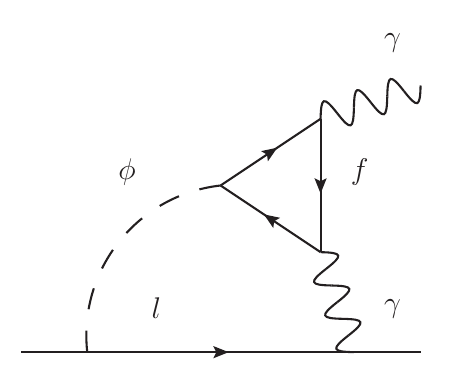}  
	\caption{The ``Barr-Zee" contribution to the $g-2$ of a lepton $l$.}
	\label{BZ}
\end{figure}
The interactions between leptons and $\phi$ described in Eq. \ref{TotLagrangian} induce new contributions to their $a_\ell$, represented in Figs. \ref{g2e} and \ref{BZ}.
The one-loop contribution of $\phi$ to $a_\ell$ in Fig. \ref{g2e} is given by  \cite{Leveille:1977rc,Tucker-Smith:2010wdq,Chen:2015vqy} 
\beq
\Delta a_\ell = \frac{\lambda_\ell^2}{8 \pi^2}\, x^2 \int_0^1 d z \frac{(1+z)(1-z)^2}{x^2(1-z)^2 + z}\,,
\label{Delaell}
\eeq
for a lepton $\ell$ of mass $m_\ell$ and $x\equiv m_\ell/m_\phi$.  The coupling of $\ell$ to $\phi$ is denoted by $\lambda_\ell$, corresponding to the interaction $\lambda_\ell \phi\bar \ell \ell$; this coupling results from the interactions in in \eq{LagrangianS}, after electroweak symmetry breaking.

The ``Barr-Zee'' diagram  contribution to $a_\ell$ (Fig. \ref{BZ}), for a heavy fermion $f$ loop, is given by \cite{Bjorken:1977vt,Barr:1990vd,Giudice:2012ms}
\beq
\Delta a_\ell^{\rm BZ}(f) = -\frac{\alpha}{6 \pi} \frac{m_\ell}{m_f} \frac{\lambda_\ell \lambda_f}{\pi^2} Q_f^2 N_c^f\, I(y)\,,
\label{2loop}
\eeq
where 
\beq
I(y) = \frac{3}{4} y^2 \int_0^1dz \frac{1 - 2 z(1-z)}{z(1-z) - y^2}\ln \frac{z(1-z)}{y^2}, 
\label{Iy}
\eeq
with $y\equiv m_f/m_\phi$; $m_f$ is the mass of $f$ and $\lambda_f$ is defined as the  coupling  $\lambda_f\phi\bar f f$.  Here, $Q_f$ and $N_c^f$ are the electric charge and the number of colors of $f$, respectively, with $N_c^f = 1(3)$ for SM leptons (quarks).  
For multiple heavy fermions $f$, one sums over them.

For fermions $f$ much heavier than the scalar, $y^2\gg 1$, the expression for $\Delta a_\ell^{\rm BZ}$ in \eq{2loop} is approximated by  \cite{Giudice:2012ms,Czarnecki:2017rlm}  
\beq
\Delta a_\ell^{\rm BZ} \approx 
\frac{\lambda_\ell\, \kappa_\gamma\, m_\ell}{4\,\pi^2} (13/12 + \ln y),
\label{DelaellBZ}
\eeq
after integrating out $f$ in the two-loop Barr-Zee diagram.  In the above, $\kappa_\gamma$ is given by  (see, {\it e.g.},  Ref.~\cite{Carena:2012xa}) 
\beq
\kappa_\gamma \approx - \frac{2\, \alpha}{3 \pi}\sum_f \frac{\lambda_f \, Q_f^2 \, N^f_c}{m_f},  
\label{kappa}
\eeq  
where the sum is over fermions with similar values of
$\ln(m_f/m_\phi)$.  In general, the terms in \eq{kappa} should be weighted by the corresponding values of the function $I(m_f/m_\phi)$.  The above formula for $\kappa_\gamma$ is obtained 
in the limit that $y^2\gg 1$.  Heavy fermions contribute to $\kappa_\gamma$ significantly only if they have sizable couplings to $\phi$.

\section{Experimental signatures}

The main experimental signatures of this model are an increase of the value of the effective Yukawa coupling between the electron and Higgs and a deviation of the electron anomalous magnetic moment. We will consider two scenarios depending on the value of the mass of the new scalar $\phi$. In the first scenario, we fix $m_\phi = 150$ GeV and let the parameters $y_e$ and $g_{S_e}$ vary in the intervals\footnote{Solutions for a negative $y_e$ are also possible with suitable choice of the other parameters.} $0<y_e<2\times 10^{-4}$ and $-1<g_{S_e}<1$,  respectively. In the second scenario, we fix  $m_\phi = 250$ GeV and make a scan over $0<y_e<5\times 10^{-5}$ and $-1<g_{S_e}<1$.   For both scenarios the other parameters are fixed to the following values:
$\mu=m_\phi$, $\lambda=1$, $\kappa= 10^{-3}$, $M_{S_e}=1.5$ TeV, $M_{e,S_e}=0$, $y_{S_e}=1$, while $\xi_e$ is determined by the condition that the mass of the electron stays at its measured value. We checked that $|\xi_e|<1$ for all the parameter space in which we are interested.

\begin{figure}[t]
	\centering
\includegraphics[width=\columnwidth]{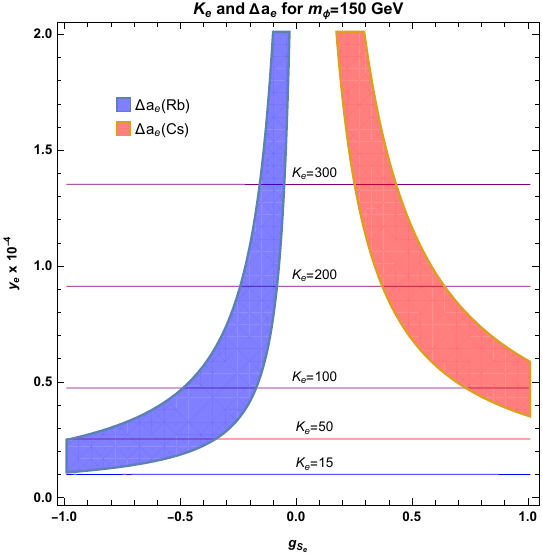}  
	\caption{Enhancement of the electron Yukawa $K_e$ [see \eq{Ke}] and $\Delta a_e$ in terms of the Lagrangian paramenters $y_e$ and $g_{S_e}$, for $m_\phi=150$ GeV.}
	\label{fig1}
\end{figure}

\begin{figure}[t]
	\centering
\includegraphics[width=\columnwidth]{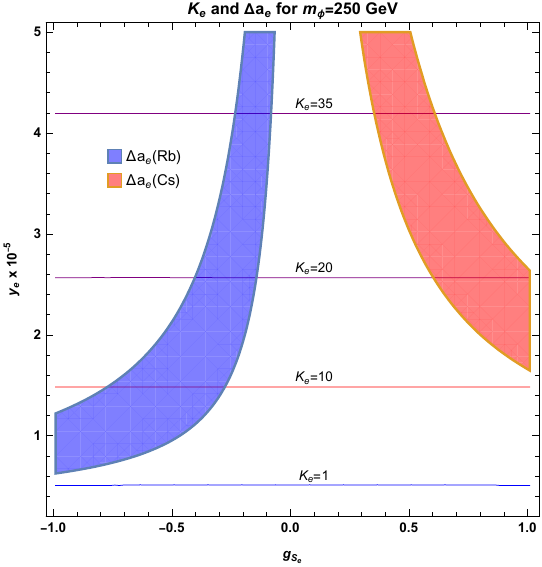}  
	\caption{Enhancement of the electron Yukawa $K_e$ [see \eq{Ke}] and $\Delta a_e$ in terms of the Lagrangian paramenters $y_e$ and $g_{S_e}$, for $m_\phi=250$ GeV.}
	\label{fig2}
\end{figure}

In Fig. \ref{fig1} and \ref{fig2} we show the results  of the aforementioned  scans for the two scenarios. The blue and red bands show the points in the parameter space that allow to generate  $\Delta a_{e}({\rm Rb})$ or $\Delta a_{e}({\rm Cs})$ respectively, within $1\sigma$ from their central values. For both scenarios the largest contribution to $a_{e}$ comes from the Barr-Zee diagram in Eq. \ref{DelaellBZ}. The horizontal lines represent different values of
\beq
K_e\equiv \left|y_e^h/y_e^{h,\rm SM}\right|\,,
\label{Ke}
\eeq
which parameterizes the enhancement of the Higgs boson coupling to the electron, compared to its SM value.    
It is interesting to notice that, in our model, a modification of the electron $a_e$ implies a larger effective Yukawa.  Moreover, the explanation of the deviation in Eq. (\ref{Del-ae-Cs-quad}) (Cs) also requires a much larger $K_e$ than the deviation in Eq. (\ref{Del-ae-Rb-quad}) (Rb). Lastly, we note that, as shown in Figs. \ref{fig1} and \ref{fig2}, increasing $g_{S_e}$ leads to smaller values of $y_e$ (and hence $K_e$) being compatible with our $1\sigma$ range for $\Delta a_e$. 
 This is because, with our choice of parameters,  the cancellation needed to get the correct electron mass, implied in \eq{EFT-me}, requires larger values of $\xi_e$, for larger values of $y_e$.  However, making $\xi_e$ larger will increase the effective  coupling of $\phi$ to electrons, which will in turn lead to $\Delta a_e$ being too large. 
 Hence, the range of $y_e$ for large $g_{S_e}$ gets limited, as illustrated in the figures. 

Regarding signatures at collider, we notice that $\phi$ could be produced through the same production mechanisms of a single Higgs. Therefore it is possible to estimate the production cross-section for $\phi$ using those of the SM Higgs. We find that the production cross sections are $\sigma(pp\to \phi)\lesssim 3$ fb and $\sigma(pp\to\phi)\ll 0.1$ fb, for $m_\phi=150$ GeV and $m_\phi=250$ GeV, respectively. Since $\phi$ decays mostly into electrons, with branching ratio $\sim93\%$ and $\sim 82\%$ for  $m_\phi=150$ GeV and $m_\phi=250$ GeV,  respectively\footnote{For $m_\phi=150$ GeV, the total width of $\phi$ is $\sim 2$ MeV, where other relevant branching ratios are $\sim 3\%$ into $Wf'\bar{f}$ and $\sim 2\%$ into $b\bar{b}$. For $m_\phi=250$ GeV, the total width $\phi$ is $\sim 28$ MeV and the other relevant branching ratios are $WW$ at $\sim 13\%$ and $ZZ$ at $\sim 5\%$.}, we can easily reinterpret the limits for $Z'$ production \cite{ATLAS:2019erb,CMS:2021ctt}. 
 We see that the direct discovery of $\phi$ seems to be beyond the capabilities of LHC, even once 3~${\rm ab}^{-1}$ of integrated luminosity is reached. On the other hand, future colliders, and particularly an $e^+ e^-$ machine would probably be able to detect $\phi$, particularly considering the current limits placed by LEP on production of new particles with large couplings to leptons \cite{Freitas:2014jla}. We have made sure that these bounds are applied in the parameter space that we have 
considered.  
Finally, regarding $S_e$, studies based on  models with similar properties to ours \cite{Bissmann:2020lge} seem to suggest that the detection of $S_e$ is also currently beyond the capabilities of the LHC.

\section{\boldmath Muon $g-2$}

There is a longstanding discrepancy between the theoretical prediction for the moun $g_\mu-2$ and its measured value.  The most recent measurements at  Fermilab \cite{Muong-2:2021ojo,Muong-2:2023cdq}, combined with a prior measurement at Brookhaven National  Laboratory \cite{Muong-2:2006rrc}, yield the experimental value 
\beq
a^{\rm exp}_\mu = 116592059(22)
\times 10^{-11}\,,
\label{amu-Exp}
\eeq
where $a_\mu\equiv (g_\mu-2)/2$.  Assuming the SM  prediction given in Ref.~\cite{Aoyama:2020ynm},
\beq
a^{\rm SM}_\mu = 116591810(43)
\times 10^{-11}\,,
\label{amu-SM}
\eeq
the discrepancy
\beq
\Delta a_\mu \equiv 
a^{\rm exp}_\mu - a^{\rm SM}_\mu = 
(249\pm 48)\times 10^{-11}
\label{Delamu}
\eeq
would have a significance of $5.2\sigma$.  We note that the status of this discrepancy is currently under scrutiny and convergence of theory around another   prediction presented in  Ref.~\cite{Borsanyi:2020mff} would reduce its significance.

By introducing  $\phi$ interactions with the muon, akin to those assumed for the electron, our setup can also provide an explanation of the deviation in \eq{Delamu}. 
For our benchmark scenario we take  $m_\phi=150$ GeV, $M_{S_e}=M_{S_\mu}=1.5$ TeV, $y_{S_\mu}=1$, $\xi_\mu=1$. With this setup, we need to impose $M_{\mu,S_{\mu}}\sim0.3$ GeV in order to guarantee that the muon mass and Yukawa coupling stay at their measured values after diagonalization. A large contribution to the muon $g-2$ comes from the Barr-Zee diagram in Eq. \ref{DelaellBZ}, which will include a sum over both $S_e$ and $S_\mu$. As a result of this, and the fact that we chose the masses of the heavy states to be the same, the Barr-Zee contribution will be proportional to $g_{S_e}+g_{S_\mu}$. We find that  $g_{S_e}+g_{S_\mu}=1$ implies
\beq
\Delta a_\mu=220 \times 10^{-11},
\eeq
which is enough to explain the deviation in Eq. \ref{Delamu}, within one $\sigma$. 

It is interesting to notice that Fig. \ref{fig1} and \ref{fig2} imply that, if $g_{S_e}+g_{S_\mu}\sim1$, only $\Delta a_e({\rm Cs})$ can be explained and either $K_e\gtrsim 80$  for $m_\phi=150$ GeV or $K_e\gtrsim 12$ for $m_\phi=250$ GeV. 
Another possible scenario is obtained by imposing $y_{S_\mu}=-1$ and $g_{S_e}+g_{S_\mu}=-1$, while keeping the other parameters the same, which yields  
\beq
\Delta a_\mu=221 \times 10^{-11}.
\eeq
In this case, Fig. \ref{fig1} and \ref{fig2} imply that only $\Delta a_e({\rm Rb})$ can be accounted for and either $K_e\gtrsim 15$  for $m_\phi=150$ GeV or $K_e\gtrsim 2$ for $m_\phi=250$ GeV.

\section*{Acknowledgements}
We would like to thank Sally Dawson for posing a question that motivated us to write this paper.  The work of H.D. is supported by the US Department of Energy under Grant Contract DE-SC0012704.  The work of P.P.G. has received financial support from Xunta de Galicia – Collaboration agreement between the Department of Culture, Education, Vocational Training, and University and the Universities of A Coruña, Santiago de Compostela, and Vigo for the reinforcement of the Research Centres of the Galician University System (CIGUS) and by European Union ERDF.

Digital data related to this work are submitted on the arXiv repository as ancillary files.
\appendix

\section*{Appendix: Flavor Symmetries}

In order to avoid bounds from flavor-changing processes in our model, one could in principle enforce lepton flavor symmetries that only allow diagonal couplings.  While the charged lepton mass matrix can be completely diagonal, this is not phenomenologically allowed for neutrinos, given the well-established  observations of neutrino oscillations.  Hence, one needs to be able to break such a symmetry.  One possibility is to assume that neutrinos have Dirac masses and that the right-handed neutrinos $\nu_R^i$, with $i=1,2,3$, are singlets of the flavor symmetry.  One can assign a separate $\mathbb Z_2$, for example, to each flavor and break them with a scalar $\chi_a$, with $a=e,\mu,\tau$.  This allows dimension-5 neutrino mass terms of the form 
\beq
\frac{\chi_a H \bar \nu_R^i L_a}{M} + \text{\small H.C.}\,, 
\label{dim5}
\eeq
up to $\ord{1}$ Wilson coefficients,  suppressed by a  scale $M$.  We take $M$ to be large, say $\sim 10^{18}$~GeV, near Planck mass.  The vev of $\chi_a$ then needs to be $\sim 10^6$~GeV for a  reasonable neutrino mass matrix. 

Note that the size of charged lepton number violating mixing mass scale allowed by this setup would be at most at the level of $\vev{\chi_a}\vev{\chi_b}/M \sim$~keV, leading to a corresponding mixing  angle $\theta \sim {\rm keV}/M_S\sim 10^{-9}$. 
 One can show that with parameters near those assumed in our preceding discussions, one could achieve sufficient suppression of flavor changing processes to avoid conflict with the data.  A detailed analysis requires more specific model building and is beyond the scope of this work.

\bibliography{el_yu}

\end{document}